\documentclass[runningheads]{llncs}
\usepackage[T1]{fontenc}
\usepackage{graphicx}
\usepackage{amsmath}
\begin{document}
\title{Velocity, Holding Time and Lifespan of Cryptocurrency in Transactions}
%
%
\author{Yu Zhang\inst{1} 
\and
Mostafa Chegeni\inst{1} \and
Claudio Tessone\inst{1,2}}
\institute{Blockchain \& Distributed Ledger Technologies, University of Zurich, Switzerland \and Blockchain Center, University of Zurich, Switzerland\\
\email{zhangyu@ifi.uzh.ch, chegeni@ifi.uzh.ch, tessone@ifi.uzh.ch}}

%
\maketitle              
\begin{abstract}
The measurement of the velocity of money is still a significant topic. In this paper, we proposed a method to calculate the velocity of money by combining the holding-time distribution and lifespan distribution. By derivation, the velocity of money equals the holding-time distribution's value at zero. When we have much holding-time data, this problem can be converted to a regression problem. After a numeric simulation, we find that the calculating accuracy is high even if we used only a small part of the holding time data, which implies a potential application in measuring the velocity of money in reality, such as digital money. We also tested the methods on Cardano and found that the method can also provide a reasonable estimation of velocity in some cases.
\end{abstract}
\section{Introduction}
The velocity of money in an economy is still a critical topic of great interest \cite{wang2003circulation} to researchers because it is tightly correlated with the nominal GDP and inflation rate based on the quantity theory of money, like the Irving Fisher equation \cite{persons1911fisher}, and serves as an indicator of an economy's health. For example, in the relationship between inflation and velocity \cite{wen2014does}, if the velocity is too high, it indicates a high inflation rate or the economy is too hot, then, a suitable monetary policy should be taken to cool down the economy. The velocity of money tends to move pro-cyclically \cite{ardakani2023dynamics,leao2005does}. Understanding the fluctuation of the velocity of money is significant to recognize the role of money in the business cycle \cite{wang2006variability}.
The velocity of money is not constant \cite{wang2006variability} and is affected by many macroeconomic factors, like the real per capita income, the nominal interest rate, and institutional changes \cite{jonung1979long,nunes2018determinants}.

How to measure the velocity of money has been a fundamental problem among the topics about economic growth. The main measuring method is based on Irving Fisher's theory, namely, the total number of money supplied ($M$) times the velocity of money ($V$) equals the price ($P$) times the gross domestic product (GDP). For example, in the Federal Reserve Economic Data (FRED), the velocity of money is calculated as the ratio of GDP to the money supply M2 that includes all of the cash people have on hand plus all of the money deposited in checking accounts, savings accounts, and other short-term saving vehicles such as certificates of deposit (CDs). However, the velocity of money in different sectors and different forms varies and is affected by many factors, like financial development \cite{akinlo2012financial}, and the aggregate method will obscure many details. 
Except for the method that is based on Irving Fisher's quantity theory, \cite{wang2003circulation} studied the statistical mechanics of money circulation in a closed economic system and found that the velocity of money is the expectation of the inverse of money's holding time. However, this method is only theoretical because all circulating money's holding time needs to be known, which is unrealistic. \cite{leontief1993money} calculated the velocity of money by the state transmission probability on an input-output system by modeling the money flow as a Markov chain process. However, a long time series of money flow is necessary for this method. 

In this paper, we proposed another method that can be promising in the calculating of velocity. The proposed method doesn't need to trace all money's holding time and it needs to trace only a small part of it. The arrangement of this paper is as follows: in Section 2, we introduce two important concepts that will be used in this paper; in Section 3, we derive the formula of the velocity of money followed by a formula check in Section 4; in Section 5, we do a numeric simulation and then test it again based on Cardano platform in Section 6; finally, we discuss in Section 7.

\section{Definition of Holding time and Lifespan}

To calculate the velocity of money, two important concepts need to be introduced in our research, namely, holding time (age) and lifespan. For any specific unit of money $b_i$, the holding time of $b_i$ is defined as the time interval from now ($t$, present time) back to its latest transaction time $\tau$, whose definition is the same as that in a working paper titled “Bitcoin Blockchain Transaction Dynamics”. Thus, $t-\tau$ is the holding time for this specific unit of money $b_i$. $\tau$ can also be interpreted as the “birth date” of the corresponding money $b_i$ and $t-\tau$ as its age. The definition of holding time is the same as the definition of age. 

At the next time step $\Delta t$, if this specific unit of money ($b_i$) is transacted, it corresponds to the “death” of this $b_i$ and its immediate rebirth, amounting to put its age counter back to zero and its lifespan (lifetime) is $t+\Delta t-\tau$, which is defined as the time interval between two consecutive transactions.
In the following section, we derive the relationship between the probability density distribution (pdf) of holding time and lifespan. After that, we calculate the velocity based on this relationship.

\section{Model Derivation of Velocity}
We define $f(x)$ as the pdf of the holding time of money whose holding time is $x$. $p(x)$ is defined as the pdf of lifespan for money whose lifespan is $x$. If we set that the present time is $t$ and the holding time (age) of a given unit of money is $x_0$, it means that the latest transaction time of this given money is $t-x_0$. 

Given the probability density (pdf) of holding time $f(x)$, we can get that the proportion of money whose age is $x_0$ is $f(x_0) \cdot dx_0$. Then in the next time step $\Delta t$, the probability that any money aging $x$ is transacted in the next time step $\Delta t$ is $\frac{p(x_0)}{S(x_0)} \cdot \Delta t$, where $S(x_0)=1-\int_{0}^{x_0} p(\tau) \, d\tau$

Thus, the proportion of money whose age is $x_0+\Delta t$ after $\Delta t$ would be $f(x_0)dx_0 \cdot (1-\frac{p(x_0)}{S(x_0)}\cdot \Delta t)$. If the process is stationary, then we get the following formula:

\begin{equation}
	\begin{cases}
		& f(x_0)dx_0 \cdot (1-\frac{p(x_0)}{S(x_0)}\cdot \Delta t)=f(x_1)dx_1  \\
		& x_1=x_0+\Delta t \\
		& dx_1=dx_0.
	\end{cases}
\end{equation}

Then, we have $f(x_0) \cdot (1-\frac{p(x_0)}{S(x_0)}\cdot \Delta t)=f(x_0+\Delta t)$. Using Taylor extension formula to extend $f(x_0+\Delta t)$, we get
\begin{equation}
	f(x_0) \cdot (1-\frac{p(x_0)}{S(x_0)}\cdot \Delta t)=f(x_0)+f(x_0)^{'} \Delta t +O(\Delta t).
	\label{eq2}
\end{equation}
After simplifying, we have the following formula: 
\begin{equation}
	\frac{p(x_0)}{S(x_0)}=-\frac{f(x_0)^{'}}{f(x_0)},
\end{equation}
where $S(x_0)=1-\int_{0}^{x_0} p(\tau) \, d\tau$.
This formula is a general conclusion that is useful in the subsequent sections. 

For money whose age is $x$, $M\cdot f(x)dx\cdot \frac{p(x)}{S(x)} \Delta t$ units of money are transacted in the next time step $\Delta t$, where $M$ is the total amount of money in circulation. Then the total number of transacted money is
\begin{equation}
	M \int_{0}^{+\infty} f(x)dx\cdot \frac{p(x)}{S(x)} \Delta t=-\Delta t \cdot M \int_{0}{+\infty} f(x) \cdot \frac{f(x)^{'}}{f(x)}dx = \Delta t \cdot M \cdot f(0).
\end{equation}

The velocity of money ($V$) is given by
\begin{equation}
	V=\frac{\Delta t \cdot M \cdot f(0)}{\Delta t \cdot M}=f(0).
\end{equation}

Now, we extend $f(x_0 + \Delta t)$ to the second order and the new equation is

\begin{equation}
	f(x_0) \cdot (1-\frac{p(x_0)}{S(x_0)}\cdot \Delta t)=f(x_0)+f(x_0)^{'} \Delta t +\frac{f(x_0)^{''}}{2}\Delta t^2+O(\Delta t^2).
\end{equation}

\begin{equation}
	\frac{p(x_0)}{S(x_0)}=-\frac{f(x_0)^{'}}{f(x_0)}-\frac{f(x_0)^{''}}{2f(x_0)}\Delta t.
\end{equation}
Then the total number of transacted money is
\begin{equation}
	\begin{split}
		M \int_{0}^{+\infty} f(x)dx\cdot \frac{p(x)}{S(x)} \Delta t=-\Delta t \cdot M \int_{0}{+\infty} f(x) \cdot (\frac{f(x_0)^{'}}{f(x_0)}+\frac{f(x_0)^{''}}{2f(x_0)}\Delta t)dx \\
		= \Delta t \cdot M \cdot f(0)+M \cdot \frac{\Delta t^2}{2}(f(\infty)^{'}-f(0)^{'}).
	\end{split}
\end{equation}
The velocity of money ($V$) is
\begin{equation}
	V=f(0)+\frac{\Delta t}{2} (f(\infty)^{'}-f(0)^{'})=f(0)-\frac{\Delta t}{2} f(0)^{'}.
\end{equation}
Here, the condition is that $f(\infty)^{'}=0$.

If we extend $f(x_0+\Delta t)$ to higher order, then the velocity ($V$) we get is:
\begin{equation}
	\begin{split}
		V=f(0)+\frac{\Delta t}{2} (f(\infty)^{'}-f(0)^{'})+\frac{\Delta t^2}{3!} (f(\infty)^{''}-f(0)^{''})+...
		+\frac{\Delta t^{(n-1)}}{n!} (f(\infty)^{(n-1)}-f(0)^{(n-1)}) \\
		=f(0)-\frac{\Delta t}{2}\cdot f(0)^{'} 
		-\frac{\Delta t^2}{3!} f(0)^{''}
		-...-\frac{\Delta t^{(n-1)}}{n!} f(0)^{(n-1)},
	\end{split}
\end{equation}
where $f(0)^{(i)}$ denotes the value of the $i^{th}$ order derivative of pdf $f(x)$ at 0. Here, the condition is that $f(\infty)^{'}=f(\infty)^{''}=f(\infty)^{'''}=...=f(\infty)^{(n-1)}=0$.

\section{Formula Checking from Holding time Distribution and Lifespan Distribution}

Firstly, we assume holding time pdf is $f(x)=\lambda \cdot e^{-\lambda x} (x>0)$ and apply formula (2) to derive the lifespan pdf $p(x)$. The process is as follows:

\begin{displaymath}
	-\frac{f(x)^{'}}{f(x)}=-\frac{-\lambda \cdot \lambda \cdot e^{-\lambda x}}{\lambda \cdot e^{-\lambda x}}=\lambda,
\end{displaymath}
namely, 
\begin{displaymath}
	\begin{cases}
		\frac{p(x)}{S(x)}=\frac{p(x)}{1-\int_{0}{x} p(\tau)d\tau}=\lambda \\
		p(x)=\lambda-\lambda \int_{0}{x} p(\tau)d\tau \\
		p(x)=\lambda e^{-\lambda x} (x>0).
	\end{cases}
\end{displaymath}

Then, the velocity is $V=f(0)=\lambda$. 

If we assume that the lifespan distribution is $p(x)=\lambda \cdot e^{-\lambda x} (x>0)$, becuase $-\frac{f(x)^{'}}{f(x)}=\frac{p(x)}{S(x)}=\frac{p(x)}{1-\int_{0}{x}p(\tau)d\tau}=\lambda$, then we can get $f(x)=\lambda \cdot e^{-\lambda x} (x>0)$. The velocity is also $V=f(0)=\lambda$.

\section{Numeric Simulation}

In this section, we do the numeric simulation to test the accuracy of our method. The economic transaction model to simulate is the same as that in \cite{wang2003circulation,dragulescu2000statistical}. We assume the economy is closed and there are $N$ agents in the economy. The total money is $M$ and each agent owns $\frac{M}{N}$ unit of money in their initial state. In each iteration, a pair of money sender and money receiver are selected randomly, and the amount of money to send is set by the formula $\Delta m = \frac{1}{2}v (m_1+m_2)$, where $m_1$ and $m_2$ is the amount of money owned by the sender and receiver, respectively. During this process, we also recorded the entropy ($E$) of the whole system and the holding time (age) of each unit of money. The entropy is defined just as \cite{wang2003circulation,dragulescu2000statistical}. For the holding time of money, they are recorded as follows: if a unit of money is transacted, then its holding time is set to zero; otherwise, its age will increase by one in each iteration step.

\begin{figure}
	\centering
	\includegraphics[width=1\textwidth]{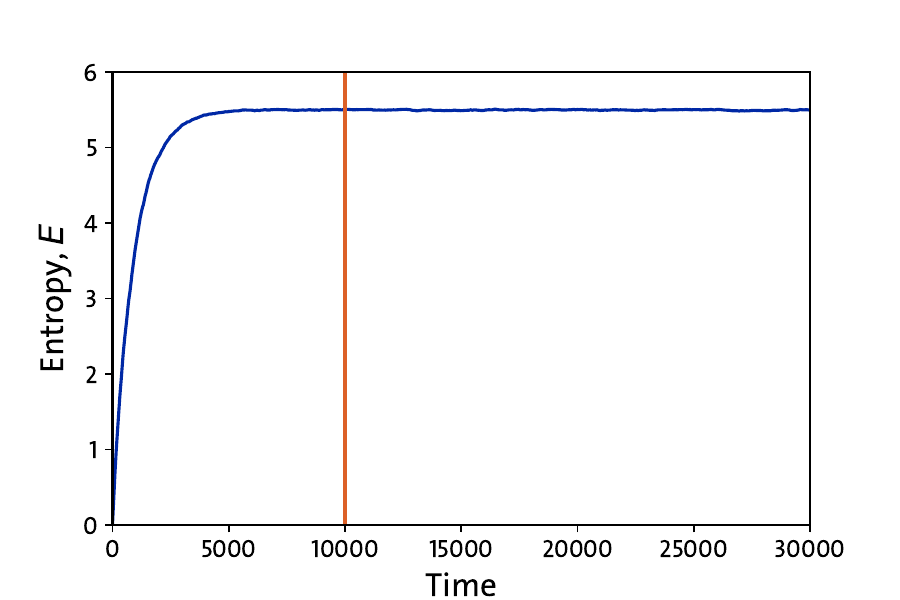}
	\caption{Entropy with time}
	\label{fig_entropy}
\end{figure}

The parameters that we set in the numeric simulation are as follows:
$m=1,000,000$ and $n=10,000$. So, the average amount of money owned by each agent is 100. As shown in Fig. \ref{fig_entropy}, we could find that the entropy of the system is almost stable after 10,000 iterations and we denote this time point as $T_0$. 

When the entropy of the system is stable,  we begin recording the transaction volume. After another 40,000 steps, we stopped the system, namely at 50,000, which is denoted as $T_1$. The data we got includes the transaction volume ($Vol$), the transacted time period ($T$, and $T=T_1-T_0$), the age of each unit of money ($age_i$, where $i$ denotes the $i^{th}$ unit of money), and the lifespan of each unit of money ($l_i$). 

The ground truth of the velocity of money can be calculated by the simple formula: $V_g = \frac{Vol}{M\cdot T}$, where $M$ is the total amount of circulating money. $V_g = 0.000134$ in this model simulation and this value will be set as a baseline to compare with our method and check whether our method is accurate in estimating the velocity of money.

\begin{figure}[!ht] 
	\centering
	\includegraphics[width=1\linewidth]{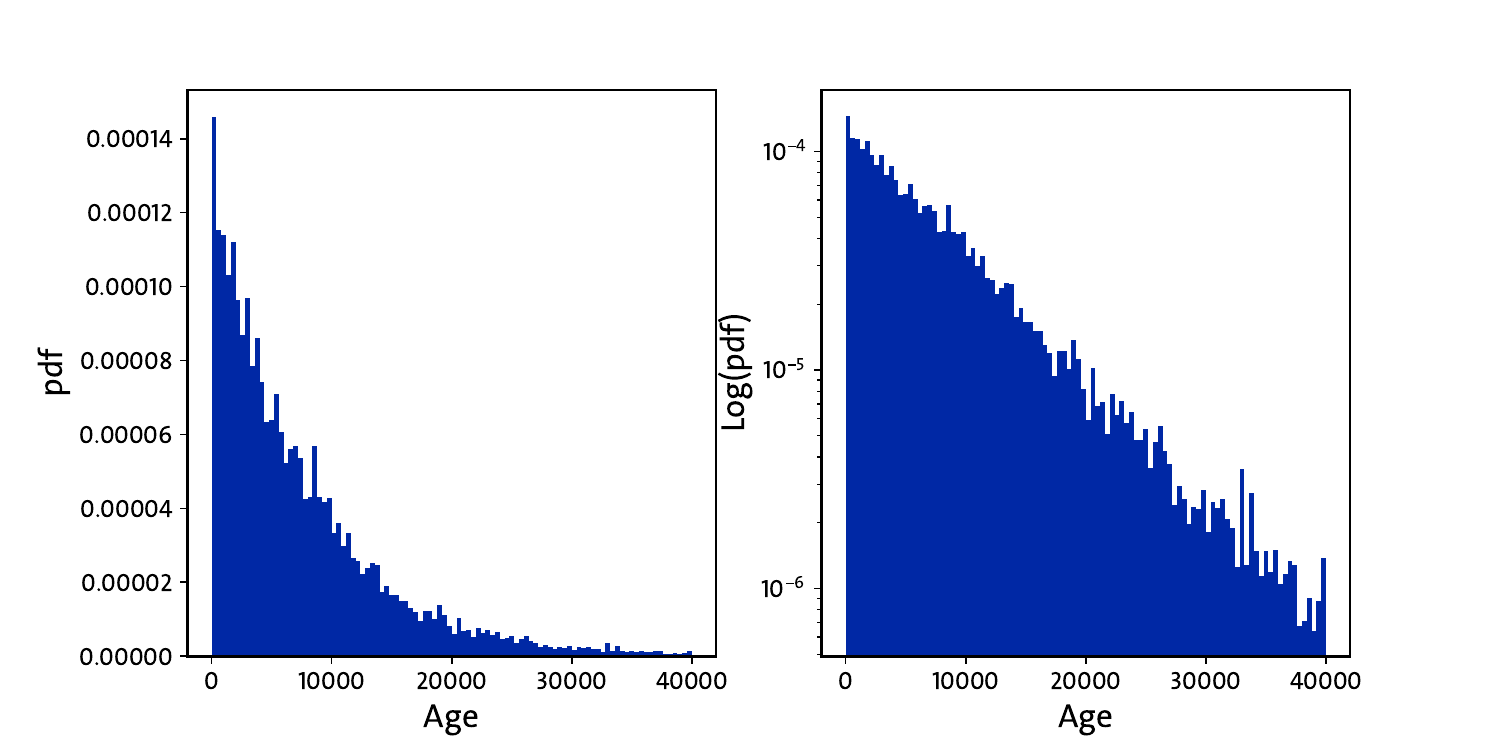}
	\includegraphics[width=1\linewidth]{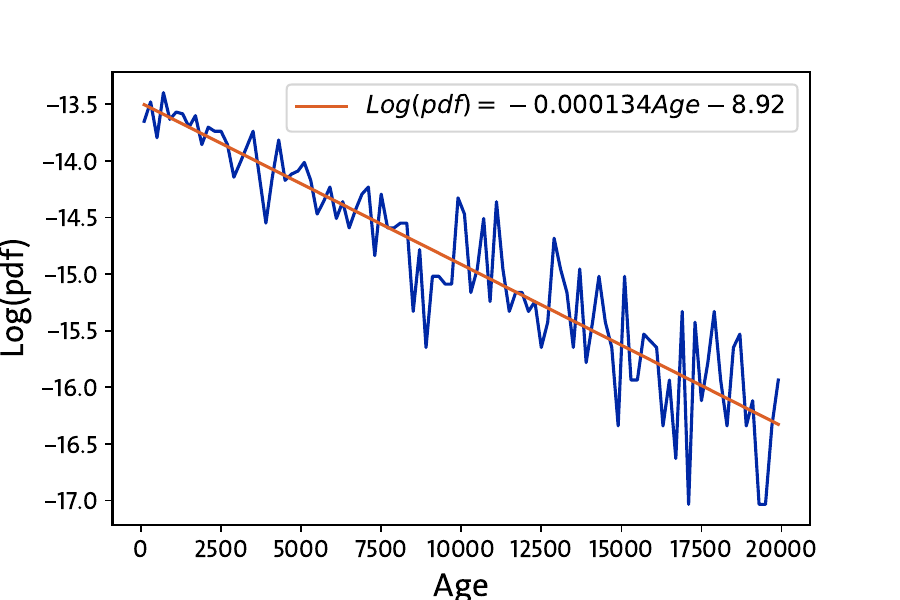}
	\caption{The age distribution of circulated money and fitting result using exponential distribution function $f(x)=\lambda e^{-\lambda x} (x \leq 0)$}
	\label{age_dist_reg}
\end{figure}

Then, we fitted the age distribution. Depending on the model, the fitted models can be different. In our model, the age distribution in theory is exponential distribution and we fitted the data based on the exponential function ($f(x)=\lambda e^{-\lambda x} (x \leq 0)$), as shown in Fig. \ref{age_dist_reg}. Based on our method, the velocity of money equals the value of the age pdf at zero, namely, $V=f(0)=\lambda$. However, the exponent of the exponential pdf is also $\lambda$. So, we can estimate the velocity of money either by calculating the value of age pdf at zero or by calculating the exponent of the function directly.

The velocity is 0.000133 by calculating the value of age pdf at zero ($f(0)$) and is 0.000134 by calculating the exponent of the exponential distribution function. We can find that they are equal or very close to the ground truth of the velocity of money ($V_g=0.000134$).

We estimated the velocity using all data based on the fitting method above. Here, a natural question is whether we can estimate the velocity using only some sample data which is always the case because it is always very hard to trace the transaction histories of all money. For example, the fiat currency we use in our daily life is physical and it is difficult to record their transaction history. In the following part, we use only partial data to fit the model and then estimate the velocity.

\begin{figure}[h]
	\centering
	\includegraphics[width=1\textwidth]{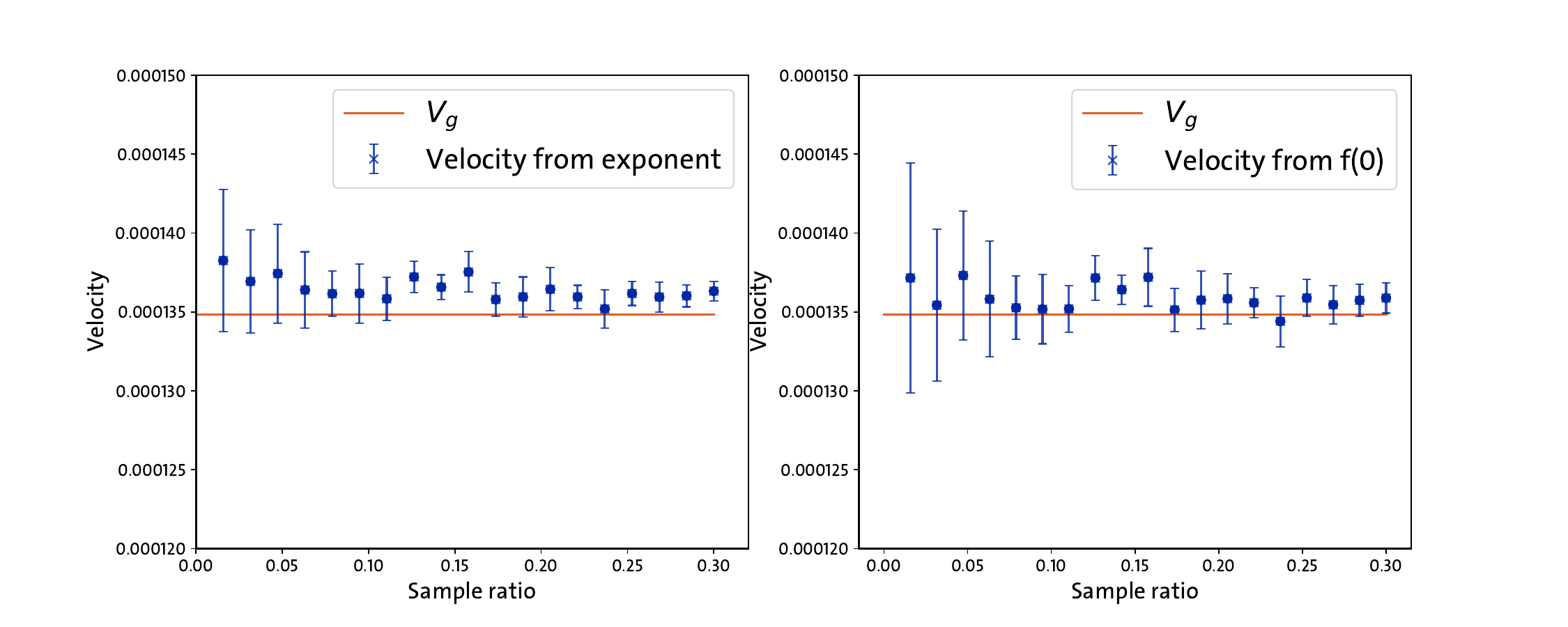}
	\caption{Measuring velocity by fitting the exponent ($\lambda$) and value at zero ($f(0)$) of holding distribution using different samples. The orange horizontal lines above both panels are the ground truth velocity. The blue dots in each bar are the mean of velocity by fitting and the length of the bar denotes the standard variation.}
	\label{sample_velocity}
\end{figure}

The ratio of our sample data ranges from 0.015 to 0.3. We sampled the holding time data and then fitted the exponent ($\lambda$) and intercept ($f(0)$) based on these sampled data to calculate the velocity of money. For each ratio, the same steps were repeated ten times to calculate the mean and the standard variation of the velocity obtained by fitting. 

As shown in Fig. \ref{sample_velocity}, we found that the more data we sampled, the smaller the variation of velocity. The velocity by fitting to sampled data is very close to the ground truth velocity ($V_g$), which denoted again that our method was effective in estimating the velocity. 

Except for fitting the holding time distribution to calculate the velocity of money, we can also fit the lifespan distribution to get the velocity, the result of which may be more accurate than fitting the holding time distribution. This is just an observation and we can not prove it in theory.

\section{Transaction Velocity in Cardano}
Cardano is a blockchain platform that uses the UTXO (Unspent Transaction Output) model, similar to other blockchains like Bitcoin. However, it goes a step further with the EUTXO (Extended UTXO) model to support more complex features like smart contracts. Smart contracts on Cardano, written in its native language Plutus \cite{chakravarty2019functional}, can execute a wide range of logic, unlike Bitcoin's simpler script capabilities. In the EUTXO system, outputs are locked with a payment address which might be linked to a payment key, a script, or a Plutus smart contract. These contracts can hold assets and release them only under specific pre-set conditions.
The backbone of Cardano's operation is its Ouroboros consensus protocol \cite{kiayias2017ouroboros}, which is based on Proof of Stake. Here, users who own ADA, Cardano's cryptocurrency, can participate in block validation and earn rewards. They can either create a new stake pool or join an existing one. Cardano's temporal structure is divided into epochs and slots, with specific pools selected algorithmically for block generation in each slot. The distribution of validation rewards is proportionate to stake contributions. 

\begin{figure}[h]
	\centering
	\includegraphics[width=0.45\textwidth]{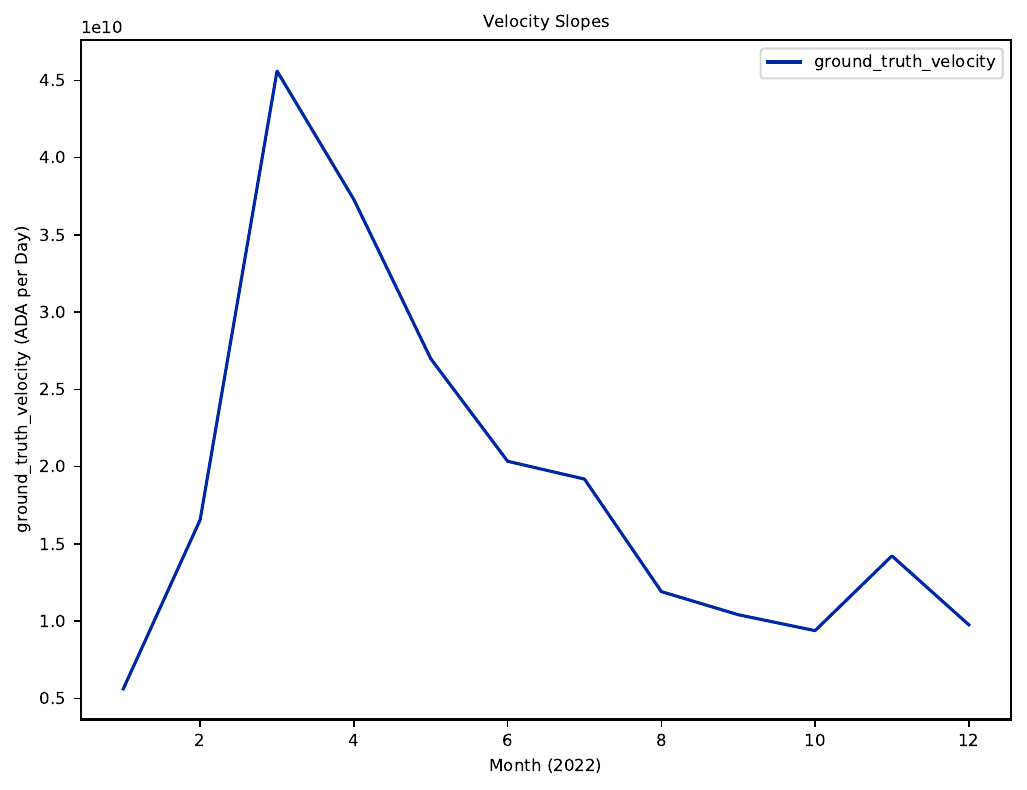}
	\includegraphics[width=0.45 \textwidth]{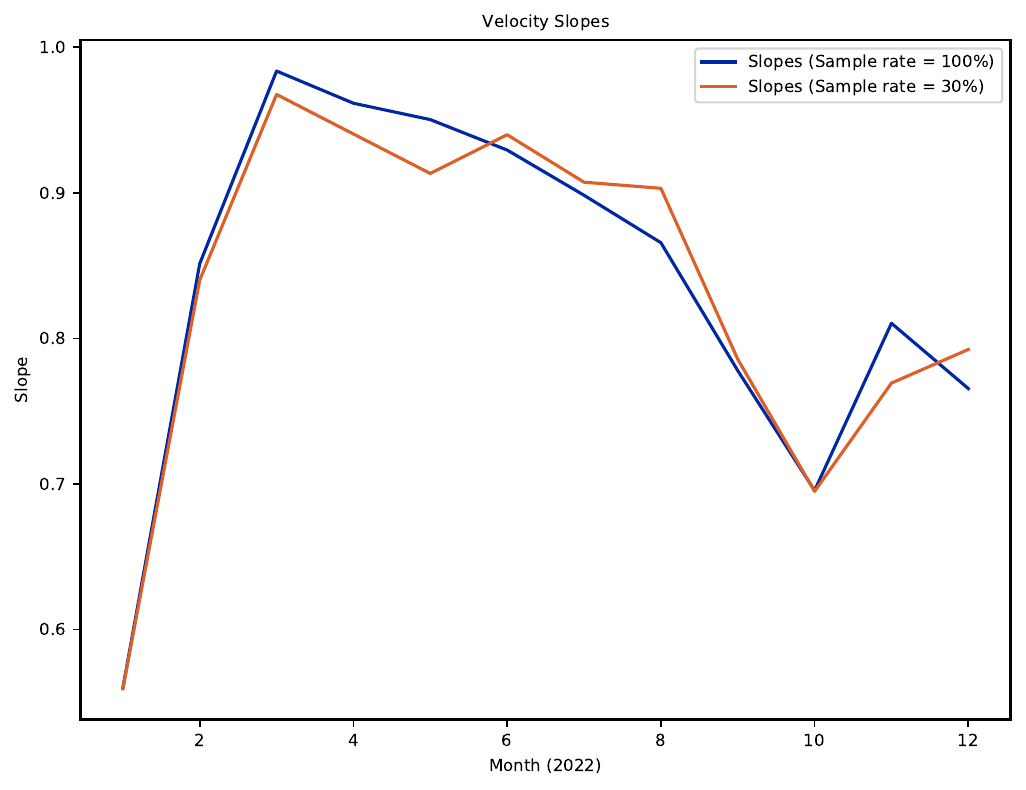}
	\caption{The left panel is the ground truth velocity and the right is the regression velocity. Note: In this pair of panels, we only compared the trends of the velocity between the ground truth and regression velocity.}
	
	\label{cardano_velocity}
\end{figure}

Just like other blockchain platforms, Cardano records each transaction on the chain. So, it facilitates us to trace the holding time for each unit of ADA. Following the steps in the numeric simulation section, we calculated the velocity of the transaction by using the holding time and compared it to the ground truth velocity. The holding time distribution is shown in Fig. \ref{fig:enter-label1} in Appendix \ref{appendix_1}. As shown in Fig. \ref{fig:enter-label1}, the holding time of ADA is fitted well by the power law function. So, for ADA, we used the power-law function to fit the holding time distribution and then used the exponent of the power-law function as the measurement of velocity. We call the velocity by fitting 'regression velocity'. In Fig. \ref{cardano_velocity}, we plotted ground truth velocity and regression velocity.

Based on the above figures,  the ground truth velocity in some months is a bit different from the regression velocity, which may be because of some large bill transactions. For example, if a transaction transacted one million ADAs one time, this large amount of transaction volume will increase the ground truth velocity a lot. Comparatively, the regression velocity will also increase, but not so much. The regression velocity can reflect the common transaction pattern change, but not some specific transactions. However, we can see that the regression velocity is similar to the ground truth velocity in trends as a whole, which denotes that our method is accurate to some extent in calculating velocity. We also calculate the regression velocities by using one hundred percent of data and thirty percent of data in the right panel in Fig. \ref{cardano_velocity}, and they are very close, which again demonstrates the usefulness of this method in estimating velocity in the case that we can get only partial data. The detailed fitting results are shown in Fig. \ref{fig:enter-label1} and \ref{fig:enter-label2} in Appendix \ref{appendix_1}.

\section{Discussion}

In this paper, we proposed a theoretical method to measure the velocity of money based on the holding time distribution and lifespan distribution. Through derivation, we ascertain that the velocity of money is equivalent to the value where the holding time pdf reaches zero when the system is stable. This result is general and it does not matter with the kinds of holding time distribution and lifespan distribution.

Can the theoretical method be applied to measure the velocity of transactions in reality? 
We tested the availability of the method in the economic system as described in \cite{wang2003circulation,dragulescu2000statistical}.
After numeric simulation, we fitted the holding time distribution data and found that the method could get a good estimation of velocity even when we have only part of the data, which means that our method may be applied in velocity estimating. We also tested the method on Cardano's transaction. To some extent, it can provide a good estimation of velocity.

It is necessary to point out that we know the holding time distribution is an exponential distribution function in the simulation model. However, it is hard to get this information in reality. Under that situation, we need to plot the holding time distribution first and then check the function that may be suitable for the data. After we get the function type, then, we can follow the steps in the numeric simulation of this paper to calculate the velocity. 
The method provided in this paper may facilitate the calculation of money velocity in reality and may be useful to policymakers and practitioners.

\bibliographystyle{unsrt}
\bibliography{reference}

\appendix

\section{Appendix}\label{appendix_1}
\begin{figure}[h]
	\centering
	\includegraphics[width=1\textwidth]{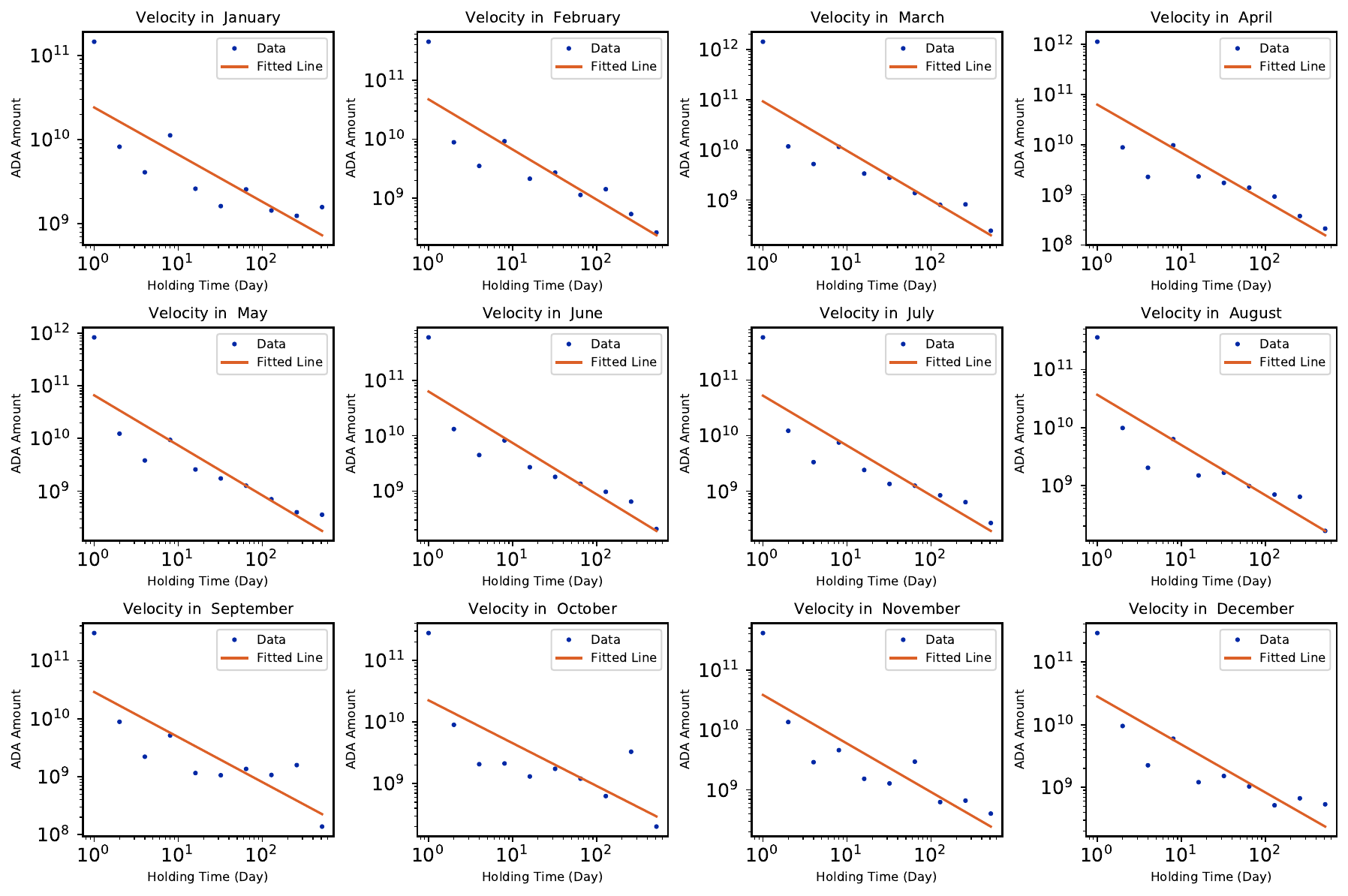}
	\caption{The holding time distribution of ADA in each month in 2022 and the fitting line based on the holding time distribution with one hundred percent of the data.}
	\label{fig:enter-label1}
\end{figure}

\begin{figure}[h]
	\centering
	\includegraphics[width=1\textwidth]{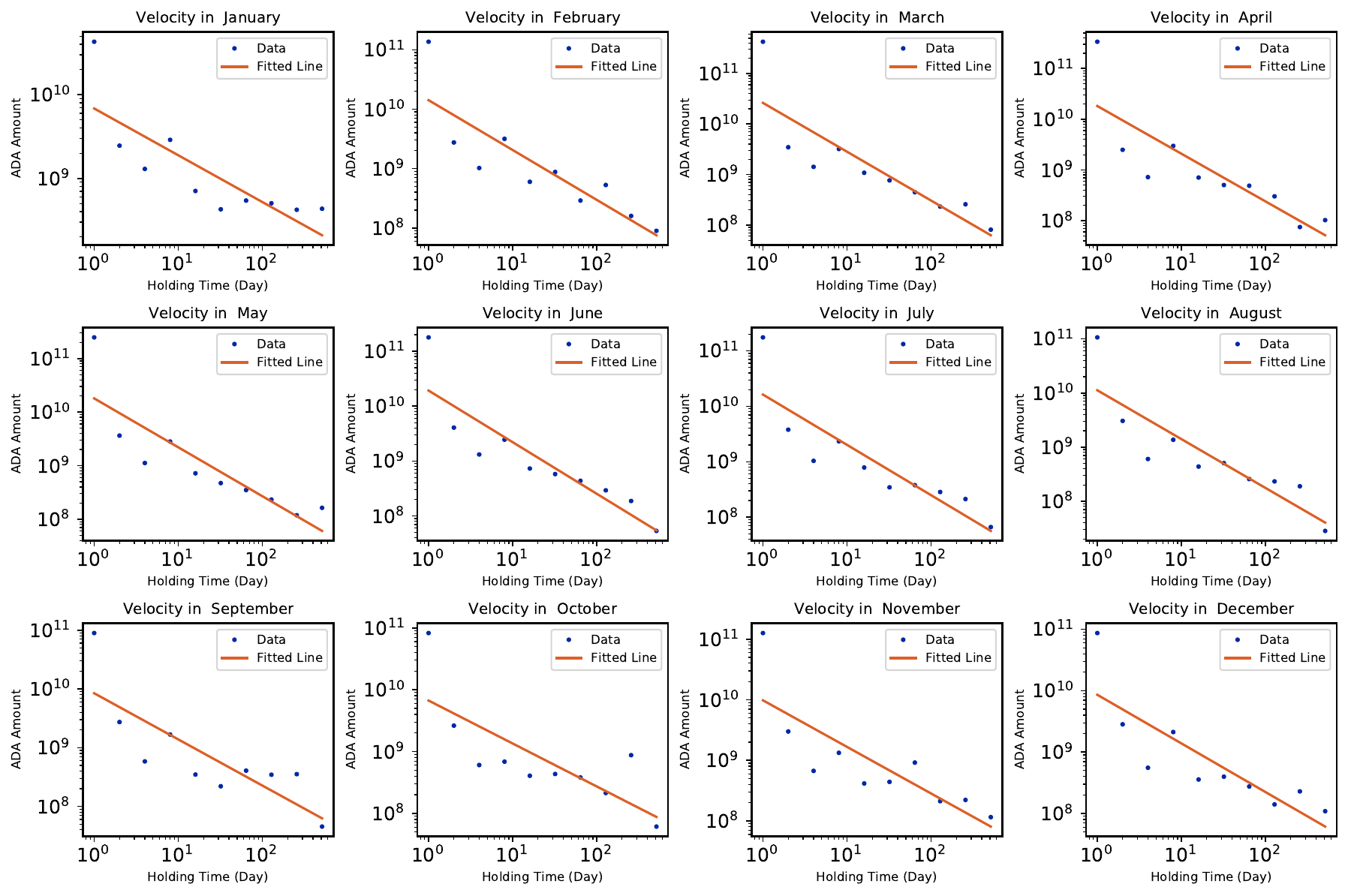}
	\caption{The holding time distribution of ADA in each month in 2022 and the fitting line based on the holding time distribution with thirty percent of data which are sampled randomly.}
	\label{fig:enter-label2}
\end{figure}

In the following figures, we plotted the holding time distribution and corresponding fitting lines. In Fig. \ref{fig:enter-label1}, we used all data, and thirty percent of data was used in Fig. \ref{fig:enter-label2}.

\end{document}